\documentclass[%
 reprint,twocolumn,
nofootinbib,
nobibnotes,
bibnotes,
 amsmath,amssymb,
 aps,
pra,
]{revtex4}
\usepackage{tcolorbox}
\usepackage[colorlinks=true, allcolors=blue]{hyperref}
\usepackage[arrowdel]{physics}
\usepackage[font=small]{caption}
\usepackage{subcaption}
\usepackage{dsfont}
\usepackage{graphicx}% Include figure files
\usepackage{dcolumn}% Align table columns on decimal point
\usepackage{bm}% bold math
\usepackage[toc,page]{appendix}

\usepackage{enumitem}
\usepackage{cleveref}

\begin{document}

\title{Effective description of a quantum particle constrained to a catenoid}

\author{G. Chac\'{o}n-Acosta}
 \email{gchacon@cua.uam.mx}
 \affiliation{Departamento de Matemáticas Aplicadas y Sistemas, Universidad Autónoma Metropolitana-Cuajimalpa, Av. Vasco de Quiroga 4871, Ciudad de México, 05348, Mexico.}%
\author{ H. Hern\'{a}ndez-Hern\'{a}ndez}%
 \email{hhernandez@uach.mx}
\affiliation{%
 Universidad Autónoma de Chihuahua, Facultad de Ingeniería, Nuevo Campus Universitario, Chihuahua 31125, Mexico.
}%
\author{ J. Ruvalcaba-Rasc\'{o}n}
\email{jose.ruvalcaba.rascon@uni-jena.de}
\affiliation{
  Abbe School of Photonics, Friedrich-Schiller-Universität  Jena, Albert-Einstein-Straße 6, 07745, Jena, Germany
}

\date{\today}

\begin{abstract}
We describe a quantum particle constrained on a catenoid, employing an effective description of quantum mechanics based on expected values of observables and quantum dispersions. We obtain semiclassical trajectories for particles, displaying general features of the quantum behaviour; most interestingly, particles present tunnelling through the throat of the catenoid, a characteristic having important physical applications. 
\end{abstract}

\keywords{Keywords: effective quantum mechanics, constrained systems, particles on surfaces}%Use showkeys class option if keyword
                              %display desired
\maketitle

\section{Introduction}

The exploration of quantum mechanics on curved surfaces has attracted the attention of the scientific community, mainly due to its potential applications across diverse fields such as condensed matter physics, quantum information processing, nanotechnology, and materials science. Recent focus has turned towards describing the intricacies of quantum mechanics on curved surfaces \cite{DASILVA201713} since, for instance, experimental techniques have synthesized various types of nanostructures with shapes resembling spheres, cylinders, and other non-trivial geometries \cite{DASILVA201713}, \cite{castro}. Among these geometries, there is one of particular interest, the catenoid, a surface of revolution obtained by rotating a catenary curve about its axis. Our work delves into the behavior of a quantum particle constrained to move on such a surface, discussing both the internal metric of the constraint manifold and the external metric of the embedding space.
Motivated by the catenoid's status as a minimal surface with unique geometric properties, our study becomes a conduit for investigating the interplay between quantum mechanics and curved spaces. We narrow our focus on the impact of constraining potentials on a particle confined to a catenoid, starting with the Schrödinger equation and discussing quantum evolution through an effective method, suited for the analysis of the system as a semiclassical dynamical one. The semiclassical evolution exhibits interesting quantum mechanical properties, displaying the profound effects of curvature on the particle's behavior. 

 The curvature of the catenoid emerges as a pivotal factor influencing the behavior of confined particles, offering a tunable parameter for manipulating electronic properties \cite{de_Lima_2021},\cite{onoe}. Additionally, the length of the catenoid becomes a controlling factor, influencing particle behavior and allowing strategic control over confinement in distinct regions of the surface \cite{szameit}. This study not only enriches our understanding of the electronic structure of curved nanotubes and graphene but also hints at novel possibilities for designing innovative nanoscale devices \cite{silva_2021,willmore,alsham}.

Starting from the standard description for a quantum particle constrained to a curved surface, given in the seminal work of da Costa \cite{daCosta81,daCosta82}, we separate the Schr\"odinger equation into normal and tangential components and constrain the particle to the surface through a confining potential acting in the normal direction \cite{daCosta81,daCosta82,teixeira_schrodinger_2019}. The analysis of the quantum evolution of the wave function has been discussed in the literature \cite{ikegami_quantum_1992,2014anphy.341..132x}, where the complexity of the dynamical equations often requires approximation methods to provide physical insight. We apply an effective description of quantum mechanics based on expectation values of observables and quantum dispersions to describe the evolution of the quantum particle on the surface \cite{bojowald-aureliano,bojowald-sandhofer}. This semiclassical implementation for the quantum evolution allows a description for individual quantum particles, whose behavior can be understood in terms of the physical parameters in the system \cite{baytas_2020,bojowald_2022,tunneling,javi}.
In the domain of semiclassical quantum mechanics on a catenoid, our effective study presents a compelling exploration into the behavior of quantum particles traversing the catenoid throat. The semiclassical particle exhibits transmission and reflection phenomena, aligning seamlessly with quantum mechanical analyses \cite{dandoloff_2009,dandoloff_2010}. Beyond theoretical insights, our findings hold potential applications in quantum information processing, enriching our understanding of quantum particle behavior in non-trivial geometries and pointing towards exciting future developments in quantum physics applications \cite{savage,mannini}. Moreover, we explore the practical implications of our findings for materials science, nanotechnology, and emerging interdisciplinary domains  \cite{silva_2021,alsham,2014anphy.341..132x,danani}.

The structure of the article is the following. In section \ref{section:curved_surface} we present the derivation of the classical and quantum description of a particle constrained to a curved surface. In section \ref{section:momentous_qm} we introduce the effective description of quantum systems, momentous quantum mechanics. In section \ref{section:semiclass_evolution} the semiclassical discussion of the quantum particle on the catenoid is presented, with the analysis of the evolution being presented in section \ref{section:numerical}. Finally, in section \ref{section:discussion} we discuss our findings, and provide possible future extensions and applications of our work.

\section{Quantum Particles on Curved Surfaces} \label{section:curved_surface}

In recent years it has been widely accepted that in dealing with the quantum mechanical problem for particles on surfaces, one has to adopt the confinement potential formalism, which predicts a geometrical potential whose physical effects can be experimentally explored \cite{de_Lima_2021,onoe,szameit,oli}. In this work, we will deal with the confinement of a particle to a two-dimensional surface embedded in a three-dimensional Euclidean space.

In this formalism \cite{daCosta81,daCosta82,teixeira_schrodinger_2019}, we take the Schr\"{o}dinger equation,
\begin{equation}
    \text{i}\hbar \frac{\partial \xi}{\partial t} = -\frac{\hbar^2}{2m} \Delta \xi + V \xi,
\end{equation}
and expressing the Laplacian in terms of the coordinates of the curved surface as in the Laplace-Beltrami operator

\begin{equation}
    \Delta f= \frac{1}{\sqrt{\abs{g}}}\sum\limits_{i,j=1}^n \partial_i \left( \sqrt{\abs{g}} g^{ij} \partial_j f\right). \label{delta}
\end{equation}

One decouples the resulting differential equation by splitting the wave function, as the product of normal ($\xi_\text{N}$) and tangential components ($\xi_\text{T}$). The most adequate parametrization of the surface is obtained by introducing a Gauss map, which yields the corresponding coordinate on the surface ($q_1,q_2$) and a normal coordinate ($q_3$). Expressing the wave function as $\xi=\xi_\text{T}(q_1,q_2) \ \xi_\text{N}(q_3)$, separates the Schrödinger equation into two decoupled equations. In the case of an interaction-free particle, we get
\begin{equation}
     \text{i}\hbar \frac{\partial \xi_\text{N}}{\partial t}= -\frac{\hbar^2}{2m}\frac{\partial^2 \xi_\text{N}}{\partial q_3^2} + V_\lambda (q_3) \xi_\text{N},\label{normal-shcrodinger-equation}
\end{equation}
and 
\begin{equation}
 \text{i}\hbar \frac{\partial \xi_\text{T}}{\partial t}=-\frac{\hbar^2}{2m}  \Delta_\text{g} \xi_\text{T} -\frac{\hbar^2}{2m}\left( H^2 -K\right)\xi_\text{T}, \label{tangential-shcrodinger-equation}
\end{equation}
where $\Delta_\text{g}$ corresponds to the operator (\ref{delta}) for the two-dimensional induced metric of the surface, and $H$ and $K$ correspond to its mean and Gaussian curvatures, respectively \cite{teixeira_schrodinger_2019}. In the first equation a confining potential is added to constraint the particle to the surface
\begin{equation}
    \lim\limits_{\lambda\rightarrow\infty} V_\lambda(q_3)=
    \begin{cases}
    0 & q_3=0,\\
    \infty & q_3\neq 0,
    \end{cases} \label{confining potential}
\end{equation}
where $\lambda$ denotes the magnitude of the binding potential. Additionally, the last term in (\ref{tangential-shcrodinger-equation}) is known as the quantum geometric potential, or da Costa potential \cite{daCosta81,daCosta82}.

\subsection{Schrödinger Equation for a Particle Constrained to a Catenoid}

Particularly, the Schrödinger equation for a particle constrained to a catenoid with an axis along $z$ and waist radius $R$ can be derived by considering the line element \cite{dandoloff_2010},
\begin{equation}
    \text{d}s^2=\cosh^2(z/R)\text{d}z^2+R^2\cosh^2(z/R)\text{d}\phi^2,
\end{equation}
from which the mean and Gaussian curvatures can be calculated, and are given respectively as
\begin{equation}
    H=0,
\end{equation}
and
\begin{equation}
    K=-\frac{1}{R^2}\sech^4\left(\frac{z}{R}\right).
\end{equation}
The geometric potential in this case is
\begin{equation}
V_G(z)=-\frac{\hbar^2}{2mR^2}\sech^4\left(\frac{z}{R}\right),\label{quantum geometric potential}
\end{equation}
which allows us to write the time-independent Schrödinger equation \cite{dandoloff_2009} in the following form

\begin{equation}
 -\frac{\hbar^2 \sech^2\left(\frac{z}{R}\right)}{2mR} \left[ R\frac{\partial^2 \xi_\text{T}}{\partial z^2 } + \frac{1}{R}\frac{\partial^2 \xi_\text{T}}{\partial \phi^2}+\sech^2 \left(\frac{z}{R}\right) \xi_\text{T} \right]=E,
\end{equation}
with $E$ the energy eigenvalue. It is important to note that the confining potential formalism predicts a non-trivial geometric potential, in this case, given by the expression (\ref{quantum geometric potential}). The geometrical nature of this potential needs to be distinguished from the characteristics of the so-called effective potential, which arises from the semiclassical approximation.

\subsection{Quantization of the Classical System}
The quantization of constrained classical systems is a long-standing problem that was earlier noticed by Dirac \cite{dirac1930,dirac2001}, and has been treated for the particular case of a particle whose motion is constrained to a curved surface. It has been shown that quantizing a particle constrained to a surface yields the Schr\"odinger equation (\ref{tangential-shcrodinger-equation}), \cite{ikegami_quantum_1992}.

To achieve this, we need to consider a conservative, constrained classical Hamiltonian written in Cartesian coordinates, obeying usual canonical relations $\left\{x_a,P_b\right\}=\delta_{ab}$, \cite{ikegami_quantum_1992},
\begin{equation}
 H=\frac{1}{2m}\sum\limits_{a,b=1}^N P_a \left(\delta_{ab} -n_a n_b\right) P_b + V(x). \label{classical}
\end{equation}
where $n_a$ are the components of the unit vector normal to the surface and $a,b=1,2,3$. Or, in terms of the surface coordinates, which are also canonical $\left\{ q_\mu, p_\nu\right\}=\delta_{\mu\nu}$, and the induced metric on the surface $g_{\mu\nu}$
\begin{equation}
    H=\frac{1}{2} g^{\mu\nu}p_\mu p_\nu.\label{constrained-classical-hamiltonian}
\end{equation}
where $\mu,\nu=1,2$.\\

When quantizing this Hamiltonian one has to define the corresponding position operators in the usual way $\hat{q}_\mu$, and the momentum operators as the geometric momentum operators defined in \cite{liu2007,pauli},
\begin{equation}
    \hat{p}_\mu= -\text{i} \hbar \left(\partial_\mu+\frac{1}{2}\Gamma_\mu \right). \label{geometrical-momentum}
\end{equation}
where $\Gamma_\mu$ corresponds to the contracted connection symbols.
 One obtains the right Schrödinger equation on the surface (\ref{tangential-shcrodinger-equation}).  The classical phase-space is coordinatized as before by $q_\mu=\langle \hat{q}_\mu\rangle$  and, $p_\mu=\langle \hat{p}_\mu\rangle$ which remain as canonical pairs. 

We emphasize that the constraint has been applied to the classical Hamiltonian, which will have no further information about the normal motion of the particle, classically there is no motion in this direction. This means that one cannot obtain (\ref{normal-shcrodinger-equation}) by quantizing this classical constrained Hamiltonian. For consistency, we will only consider the extreme case for the confining potential given in (\ref{confining potential}).

\section{Semiclassical description of quantum mechanics} \label{section:momentous_qm}

Momentous quantum mechanics describes effectively the quantum dynamics via the Taylor expansion of the expectation value of the quantum Hamiltonian operator. This expansion leads to a quantum-corrected semiclassical Hamiltonian that yields enhanced classical equations of motion and introduces dynamical dispersions called quantum variables \cite{bojowald-aureliano}.

The quantum variables are defined as
\begin{align}
    G^{a_1,b_1,\cdots,a_k,b_k}&\equiv& \nonumber\\ \langle \left(\hat{q}_1-q_1 \right)^{a_1}(\hat{p}_1 -p_1)^{b_1}&\cdots&\left( \hat{q}_k-q_k\right)^{a_k}(\hat{p}_k-p_k )^{b_k} \rangle ,
    \label{definition-of-moments}
\end{align}
%
%The quantum variables are defined as
%\begin{widetext}
%\begin{equation}
%    G^{a_1,b_1,\cdots,a_k,b_k}\equiv \langle \left(\hat{q}_1-q_1 \right)^{a_1}(\hat{p}_1 -p_1)^{b_1}\cdots\left( \hat{q}_k-q_k\right)^{a_k}(\hat{p}_k-p_k )^{b_k}\rangle _\text{Weyl}, \label{definition-of-moments}
%\end{equation}
%\end{widetext}
%
where $k$ corresponds to the number of degrees of freedom of the system, and for any of the operator products, from now on, we will consider the Weyl ordering. Expectation values of canonical operators $q_k=\langle\hat{q}_k\rangle$ and $p_k=\langle\hat{p}_k\rangle$, satisfying $\left[\hat{q},\hat{p} \right]=\text{i} \hbar\mathds{1}$, now corresponding to the classical variables $x_k$, and $p_k$, coordinatize the augmented phase space with the $G^{a_1,b_1,\cdots,a_k,b_k}$ variables \cite{bojowald-sandhofer}. %Expectation values of canonical operators $q_k=\langle\hat{q}_k\rangle$ and $p_k=\langle\hat{p}_k\rangle$, satisfying $\left[\hat{q},\hat{p} \right]=\text{i} \hbar\mathds{1}$,  coordinatize the augmented phase space, now corresponding to classical variables $x_k$, and $p_k$ \ref{bojowald-sandhofer}.

The symplectic structure of phase space determines a Poisson bracket \cite{bojowald-aureliano} algebra for the expectation values,
\begin{equation}
    \left\{\langle\hat{A}\rangle,\langle \hat{B}\rangle   \right\} = \frac{1}{\text{i}\hbar} \Big\langle \left[\hat{A},\hat{B}\right]\Big\rangle .
\end{equation}
 We can see that classical variables are canonical. The Poisson brackets of the quantum variables are given in Appendix \ref{appendix-A}.
 
 Momentous quantum mechanics is based on a quantum-corrected Hamiltonian, which is defined as the expectation value of the quantum Hamiltonian operator in the following manner,
\begin{widetext}
\begin{equation}
    \langle\hat{H}\rangle\equiv H_\text{Q}=H\left(\{x_i\},\{P_i\}\right)+\sum\limits_{a_1,b_1}^\infty \cdots \sum\limits_{a_k,b_k}^\infty \frac{\partial^{a_1+b_1+\cdots+a_k+b_k} H}{\partial x_1^{a_1}\partial P_1^{b_1}\cdots\partial x_k^{a_k}\partial P_k^{b_k}} \frac{G^{a_1,b_1;\cdots;a_k,b_k} }{a_1!b_1!\cdots a_k!b_k!}, \label{definition-of-qhamiltonian}
\end{equation}
\end{widetext}%
where $H$ is the classical Hamiltonian of the system and $\hat{H}$ is the quantum Hamiltonian operator. In this paper, we refer to the quantum-corrected Hamiltonian $H_\text{Q}$ simply as the quantum Hamiltonian, and should not be confused with the quantum Hamiltonian operator $\hat{H}$, which will not be used unless explicitly specified. 

The symplectic structure allows us to calculate the equations of motion of the variables as
\begin{equation}
    \dot{f}=\left\{ f,H_\text{Q}\right\}, \label{qh}
\end{equation}
where $f$ can be either a classical or a quantum variable. Observe that, starting from the Poisson brackets between classical and quantum variables (\ref{GX}) and (\ref{GG}), quantum variables only {\it see} other quantum variables, here the interaction between classical and quantum variables is given by the potential of the quantum Hamiltonian.

The effective description inherits Heisenberg's uncertainty relations, which are now written in terms of the dispersions, as
\begin{eqnarray}
    G^{2,0,0,0} G^{0,2,0,0}-(G^{1,1,0,0})^2&\geq& \frac{\hbar^2}{4},\label{uncertainty-for-theta}\\
    G^{0,0,2,0} G^{0,0,0,2}-(G^{0,0,1,1})^2&\geq& \frac{\hbar^2}{4}, \label{uncertainty-for-z}
\end{eqnarray}
for a system with two degrees of freedom.

In momentous quantum mechanics, it is required that the extended classical Hamiltonian (\ref{definition-of-qhamiltonian}) must be a conservatively constrained one. Additionally, the momentum operators in the definition of the quantum variables must be the geometric momenta as defined in (\ref{geometrical-momentum}). These requirements are necessary for our description to agree with the full quantization scheme for the constrained classical particle. 

It is worth mentioning that a new complete and canonical formulation of momentous quantum mechanics has been recently introduced \cite{baytas_2020,bojowald_2022}. However, the characteristics of the restricted classical Hamiltonian that we will consider restrict the use of this formulation, as can be followed in Appendix \ref{appendix-B}. Given this fact, we use the formulation of momentous quantum mechanics presented above for the following sections.

\section{Semiclassical Description for the Particle on a Catenoid Problem} \label{section:semiclass_evolution}

We consider the conservative classical Hamiltonian describing the motion of a particle constrained on a  catenoid with fixed $R$ in cylindrical coordinates, whose Hamiltonian is \cite{2014anphy.341..132x},
\begin{equation}
    H =\frac{ \sech^2\left(z/R\right) }{2mR^2} \left( p_\theta^2 +R^2p_\text{z}^2 \right). \label{classical-hamiltonian}
\end{equation}
We use the definition for the geometric momentum operators (\ref{geometrical-momentum}) for the moments, which in this particular case are
\begin{eqnarray}
    \hat{P}_\theta&=&-\text{i}\hbar \partial_\theta,\label{theta-momentum}\\
    \hat{P}_\text{z} &=& -\text{i} \hbar \left[\partial_\text{z}+\frac{1}{R}\tanh\left(\frac{z}{R}\right)\right].\label{z-momentum}
\end{eqnarray}

The quantum moments using (\ref{definition-of-moments}), with canonical coordinates $\theta$ and $z$ are defined as
\begin{equation}
    G^{a,b,c,d}\equiv \langle (\hat{\theta}-\theta_0)^{a}(\hat{P}_\theta -P_{\theta_0})^{b}\left( \hat{z}-z_0\right)^{c}(\hat{P}_\text{z}-P_{z_0} )^{d}\rangle, %_\text{Weyl},
\end{equation}
where the product of operators is symmetrical Weyl ordered. The quantum effective Hamiltonian obtained from (\ref{classical-hamiltonian}), using (\ref{definition-of-qhamiltonian}) is
\begin{eqnarray} \label{effective_Hamiltonian}
    H_\text{Q}&=& H+ \frac{1}{2mR^2} \sum\limits_{b+c+d\geq2}^\infty \sech^2 \left(\frac{z}{R}\right) \text{P}_c\left(\tanh\left(\frac{z}{R}\right) \right)\nonumber\\
    &\times& \frac{ G^{0,b,c,d}}{b!c!d!}\left[\prod\limits_{j=0}^{b-1} (2-j) P_\theta^{2-b}+R^2 \prod\limits_{k=0}^{d-1} (2-k) P_\text{z}^{2-d}\right],\nonumber\\
    ~
\end{eqnarray}
where $0\leq b\leq 2$,  $0\leq d \leq 2$ and $\text{P}_n(\tanh x)$ is the polynomial \cite{boyadzhiev2010derivative,Feng12017tan},
\begin{equation}
    \text{P}_n\left(\tanh x \right) = -2^{n+1} \sum\limits_{k=1}^{n+1} \frac{k!}{2^k}\mathcal{S}_{n+1}^{(k)} \left( \tanh x -1 \right)^{k-1},
\end{equation}
where $\mathcal{S}_n^{(k)}$ are the Stirling numbers of second kind.

\subsection{Second Order Truncation}

The dynamical system obtained from (\ref{effective_Hamiltonian}) comprises the equations of motion of the classical variables and the infinite many quantum variables. Evidently, solving such a system analytically is impossible, so it is necessary to perform consistent truncations.

Therefore, up to the second order, we write this Hamiltonian simply as,
\begin{eqnarray} \label{second_order_Hamiltonian }
    H_\text{Q} &=& \frac{\sech^2\left(z/R\right) }{2mR^2} \bigg\{p_\theta^2 +R^2p_\text{z}^2 -\frac{4}{R}\tanh\left(\frac{z}{R}\right) p_\theta^2 G^{0,1,1,0}\nonumber\\
    &-&4R\tanh\left(\frac{z}{R}\right) p_\text{z} G^{0,0,1,1} +2 G^{0,2,0,0} +2R^2G^{0,0,0,2} \nonumber\\
    &+&\frac{2}{R^2} \left[2-3\sech^2\left(\frac{z}{R}\right)\right]\left(p_\theta^2+R^2p_\text{z}^2\right)G^{0,0,2,0}\bigg\}. \label{approx-hamiltonian}
\end{eqnarray}
From the Poisson brackets with the effective Hamiltonian (\ref{approx-hamiltonian}), and employing the relation (\ref{GX}), the system of equations of motion for the classical variables is
\begin{eqnarray}
    \dot{\theta} &=& \frac{\sech^2(z/R)}{mR^2} \bigg\{ p_\theta - \frac{4}{R}\tanh\left(\frac{z}{R}\right)p_\theta G^{0,1,1,0} \nonumber \\
    & +&\frac{2p_\theta}{R^2}\left[2-3\sech^2\left(\frac{z}{R}\right] \right)G^{0,0,2,0}\bigg\}, \nonumber\\
    \dot{p}_\theta &=& 0, \nonumber\\
    \dot{z} &=& \frac{\sech^2(z/R)}{mR^2} \bigg\{ R p_\text{z} - 2\tanh\left(\frac{z}{R}\right) G^{0,0,1,1} \nonumber \\
    &+&\frac{2p_\text{z}}{R}\left[2-3\sech^2\left(\frac{z}{R}\right)\right]G^{0,0,2,0}\bigg\}, \nonumber\\
    \dot{p}_\text{z} &=& \frac{2}{R}\tanh(\frac{z}{R})H_\text{Q} - \frac{2}{mR^2}\sech^4\left(\frac{z}{R}\right) \nonumber \\
    &\times& \bigg[ \frac{3}{R^3}\left(p_\theta^2+R^2 p_\text{z}^2\right)\tanh(\frac{z}{R})G^{0,0,2,0} \nonumber\\
    &-& p_\text{z} G^{0,0,1,1}- \frac{p_\theta^2}{R^2}G^{0,1,1,0}\bigg].
\end{eqnarray}
Given the symplectic structure in (\ref{GG}), and defining the common factor $\eta(z)=\sech^2\left(z/R\right)/2mR^2$ we have the system of equations of motion for the quantum variables as
\begin{eqnarray*}
    \Dot{G}^{1,1,0,0}&=& 4\eta(z) \left[-\frac{\tanh\left(z/R\right)}{R}p_\theta^2G^{0,1,1,0}+G^{0,2,0,0} \right],\nonumber\\
    \Dot{G}^{1,0,1,0}&=& 4\eta(z) \bigg[-\frac{\tanh\left(z/R\right)}{R}p_\theta^2G^{0,0,2,0}+R^2 G^{1,0,0,1} \\
    &+&G^{0,1,1,0}-R\tanh\left(\frac{z}{R}\right) p_\text{z} G^{1,0,1,0}  \bigg],\\
    \Dot{G}^{1,0,0,1}&=&4\eta(z)\bigg\{-\frac{\tanh\left(z/R\right)}{R}p_\theta^2\left(G^{0,0,1,1}-G^{1,1,0,0}\right)\\ 
    &+&R\tanh\left(\frac{z}{R}\right)p_\text{z} G^{1,0,0,1}+G^{0,1,0,1}\\
    &-&\frac{(p_\theta^2+R^2p_\text{z}^2)}{R^2} \left[2-3\sech^2\left(\frac{z}{R}\right) \right] G^{1,0,1,0} \bigg\},\\
     \Dot{G}^{0,1,1,0}&=& 4R\eta(z) \left[-\tanh\left(\frac{z}{R}\right)p_\text{z}^2G^{0,1,1,0} +R G^{0,1,0,1} \right],\\
      \Dot{G}^{0,1,0,1}&=&\frac{4}{R}\eta(z) \bigg\{\tanh\left(\frac{z}{R}\right)\left[p_\theta^2G^{0,2,0,0} +R^2p_\text{z} G^{0,1,0,1}\right]\\
       &-&\frac{(p_\theta^2+R^2p_\text{z}^2)}{R} \left[2-3\sech^2\left(\frac{z}{R}\right) \right] G^{0,1,1,0} \bigg\},
    \end{eqnarray*}
     \begin{eqnarray}
     \Dot{G}^{0,0,1,1}&=& \eta(z) \bigg\{\frac{4\tanh\left(z/R\right)}{R}p_\theta^2G^{0,1,1,0}+4R^2 G^{0,0,0,2}\nonumber\\
     &-&\frac{4(p_\theta^2+R^2p_\text{z}^2)}{R^2} \left[2-3\sech^2\left(\frac{z}{R}\right) \right] G^{0,0,2,0}\bigg\},\nonumber\\
    \Dot{G}^{2,0,0,0}&=&\eta(z) \left[-\frac{4\tanh\left(z/R\right)}{R}p_\theta^2G^{1,0,1,0} +8G^{1,1,0,0}\right],\nonumber\\
         \Dot{G}^{0,2,0,0}&=& 0,\nonumber\\
  \Dot{G}^{0,0,2,0}&=& \eta(z) \left[-8R\tanh\left(\frac{z}{R}\right) p_\text{z} G^{0,0,2,0} +8R^2 G^{0,0,1,1}\right],\nonumber\\
  \Dot{G}^{0,0,0,2}&=& \eta(z) \bigg\{\frac{8\tanh\left(z/R\right)}{R}p_\theta^2G^{0,1,0,1}\nonumber\\   &-&\frac{8(p_\theta^2+R^2p_\text{z}^2)}{R^2} \left[2-3\sech^2\left(\frac{z}{R}\right) \right] G^{0,0,1,1}\nonumber\\
  &+&8R\tanh\left(\frac{z}{R}\right) p_\text{z} G^{0,0,0,2} \bigg\}
 .
     \end{eqnarray}

To find the initial conditions to solve the equations of motion, we propose a normalized Gaussian wave packet as our initial state,
\begin{eqnarray}
\psi_0(\theta,z)&=&\left[\frac{\lambda^{1/2}}{R\pi \sigma_\text{z} \erf(\pi\lambda^{1/2})}\right]^{1/2}\nonumber\\
&\times&\exp{\text{i}(lz+m\theta)-\frac{(z-z_0)^2}{2\sigma_\text{z}^2}-\frac{\lambda (\theta-\theta_0)^2}{2}},\nonumber\\
~
\end{eqnarray}
where $-\infty<z<\infty$ and $-\pi \leq \theta < \pi$ and $\lambda$ and $\sigma_z$ are related to the initial width of dispersion in $\theta$ and $z$, respectively. Using this state, we find the initial values of the quantum variables by evaluating the respective expected values in (\ref{definition-of-moments}) and using the generalized momentum operators in (\ref{theta-momentum}-\ref{z-momentum}). We get 
\begin{eqnarray}
    ~G_0^{1,1,0,0}&=&~G_0^{1,0,1,0}=  G_0^{1,0,0,1}= G_0^{0,1,1,0}\nonumber\\
    &=&G_0^{0,1,0,1}= G_0^{0,0,1,1}=0,\nonumber\\
    ~ G_0^{2,0,0,0}&=& \frac{1}{2\lambda\erf(\pi \lambda^{1/2})} \bigg(\erf\left(\pi \lambda^{1/2}\right)\nonumber\\
    &-&(\pi\lambda)^{1/2}\exp{-\pi^2\lambda} \bigg),\nonumber\\
    ~ G_0^{0,2,0,0}&=&\hbar^2 \lambda \left(1-\lambda G_0^{2,0,0,0}\right),\nonumber\\
    ~ G_0^{0,0,2,0}&=&\frac{1}{4}\sigma^2 \left(2+\frac{\sigma^2}{R^2} \right).
\end{eqnarray}
The initial value for $G^{0,0,0,2}$ does not have an analytical expression, thus its value was calculated by numerical integration. We point out that although the initial values for the moments were calculated for a specific initial wave function, the dynamical evolution for these quantum variables does not depend on a specific state. The system of equations of motion for the moments comes from the effective Hamiltonian. 

This effective setting allows exploring general quantum effects, such as quantum tunneling \cite{tunneling}, that for certain states for the analysis of the catenoid may not appear. We will see that this is the case here as we report the reflection and transmission of particles through semiclassical trajectories. In previous studies, as in \cite{dandoloff_2010}, no reflection has been found due to the choice of states. We will discuss this later on.

\section{Numerical evolution} \label{section:numerical}

Numerically, we will use $\hbar=m=1$, $R=1/2$, and the $z$-momentum will be taken as a varying parameter $-a$. We employ the following initial numerical values of the classical and quantum variables
\begin{equation}
\begin{split}
\theta_0 &=0,\\
 P_{\theta_0} &= 1,\\
 z_0 &= 1\\
 P_{z_0} &= -a,\\
 G_0^{1,1,0,0} &= 0,\\
 G_0^{1,0,1,0} &= 0,\\
 G_0^{1,0,0,1} &= 0,
\end{split}
\hspace{1cm}
\begin{split}
G_0^{0,1,1,0} &= 0,\\
 G_0^{0,1,0,1} &= 0,\\
 G_0^{0,0,1,1} &= 0,\\
 G_0^{2,0,0,0} &= 0.05,\\
 G_0^{0,2,0,0} &= 5,\\
 G_0^{0,0,2,0} &= 0.06,\\
 G_0^{0,0,0,2} &= 4.169094014469417.
\end{split}
\end{equation}

With these initial conditions, one can solve the system of equations numerically since no analytical solution exists. We now explore the behavior of the resulting quantum-corrected trajectories.

\subsection{Single Particle Effective Trajectories}

\begin{figure}
\centering
\includegraphics[width=1.1\linewidth]{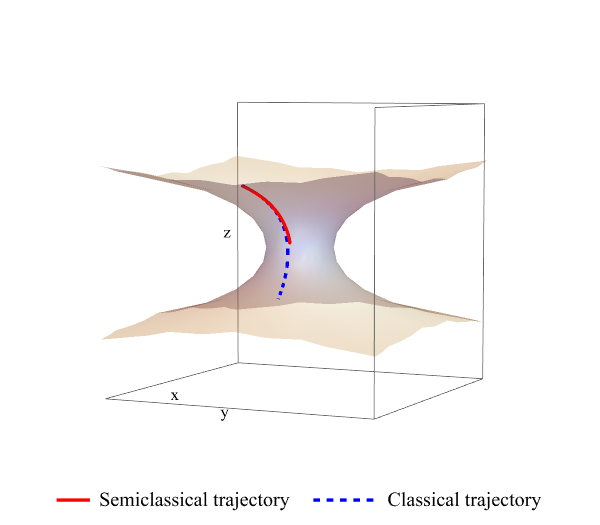}
\caption{Semiclassical and classical trajectories for a particle on a catenoid with $a=1$.}
\label{free-single-cvsq-trajectory}
\end{figure}

Let us discuss the case of a single particle by fixing $a=1$ for the classical and effective solutions. Figure \ref{free-single-cvsq-trajectory} compares both trajectories on the surface of the catenoid for a range of $0\leq t \leq 0.25$. Notice that the particle starts on the upper half of the surface and, while the classical trajectory continues to the lower half, the semiclassical presents a rebound on the throat, confining its motion to $z>0$. This effect is due entirely to the quantum nature of the system.  

\begin{figure}
\centering
\begin{subfigure}{0.35\textwidth}
  \centering
\includegraphics[width=1\linewidth]{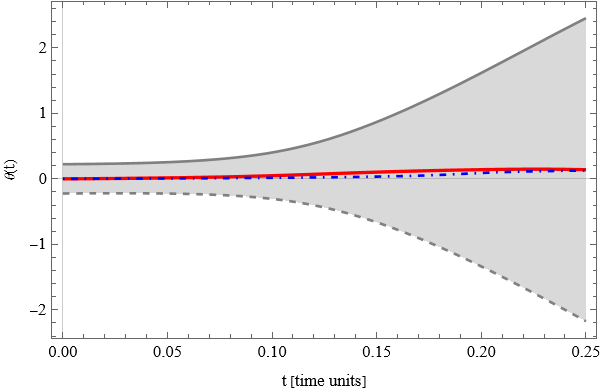}
\caption{$\theta$ evolution.}
\label{free-single-cvsq-theta}
\end{subfigure}%
  \hfill
\begin{subfigure}{0.35\textwidth}
  \centering
\includegraphics[width=\linewidth]{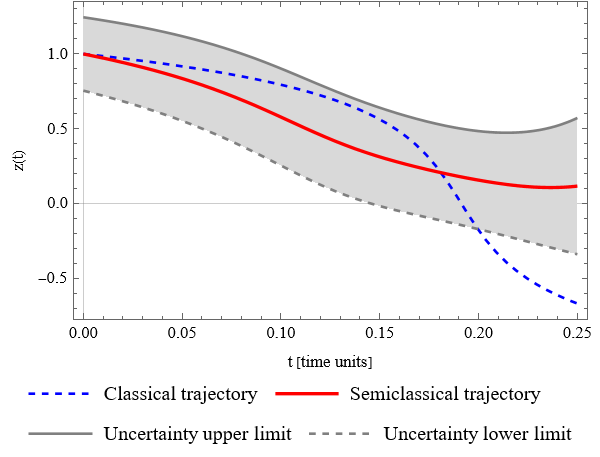}
\caption{$z$ evolution.}
\label{free-single-cvsq-z}
\end{subfigure}
\caption{Variable evolution for the semiclassical and classical cases with $a=1$, where the uncertainty belt (in grey) is shown.}
\label{fig:test}
\end{figure}

Referring to Figure \ref{free-single-cvsq-theta}, we can see that the trajectories do not vary much in the angular variable since the quantum correction for $\theta$ is small. In Figure \ref{free-single-cvsq-z}, we observe that for the $z$-variable, the trajectories acquire considerable quantum corrections, the classical particle goes through the catenoid throat, while the quantum-corrected one does not reach $z=0$. We also show the semiclassical trajectory's lower and upper uncertainty bounds, making its probabilistic nature explicit. As we mentioned above, the semiclassical trajectory has a bounce.

\begin{figure}
\centering
\begin{subfigure}{.35\textwidth}
  \centering
\includegraphics[width=1\linewidth]{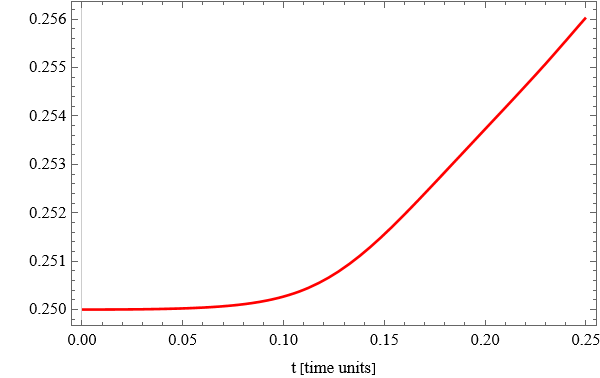}
\caption{$ G^{2,0,0,0} G^{0,2,0,0}-(G^{1,1,0,0})^2\geq \hbar^2/4$.}
\label{free-single-theta-uncertainty}
\end{subfigure}%
  \hfill
\begin{subfigure}{.35\textwidth}
  \centering
\includegraphics[width=1\linewidth]{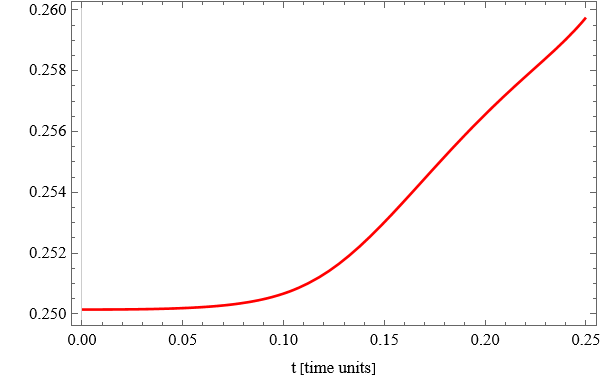}
\caption{$G^{0,0,2,0} G^{0,0,0,2}-(G^{0,0,1,1})^2\geq \hbar^2/4$.}
\label{free-single-z-uncertainty}
\end{subfigure}
\caption{Uncertainty relations for the variables with $a=1$.}
\label{fig:test2}
\end{figure}

As a validity condition, we verify that the uncertainty relations for the moments in (\ref{uncertainty-for-theta}) and (\ref{uncertainty-for-z}) are satisfied. In Figures \ref{free-single-theta-uncertainty} and \ref{free-single-z-uncertainty}  we show that this is the case.

\subsection{Multiple Quantum-Corrected Trajectories}

\begin{figure}[h] 
\centering
\begin{subfigure}[b]{0.45\textwidth}
 \centering
\includegraphics[width=0.8\linewidth]{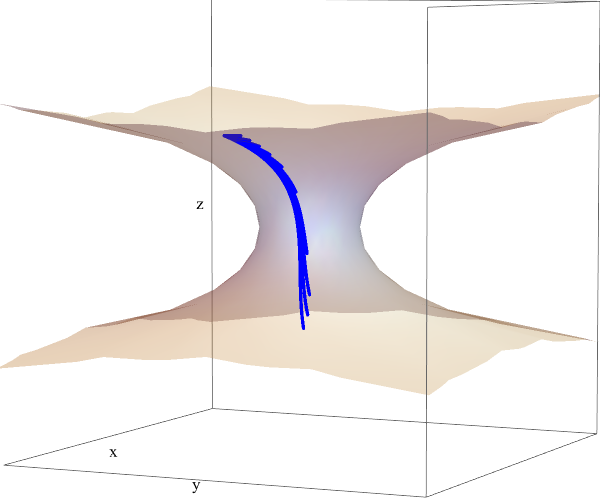}
\caption{Classical trajectories.}
\label{free-multiple-classical-trajectories}
\end{subfigure}%
  \hfill
\begin{subfigure}[b]{0.45\textwidth}
  \centering
\includegraphics[width=0.8\linewidth]{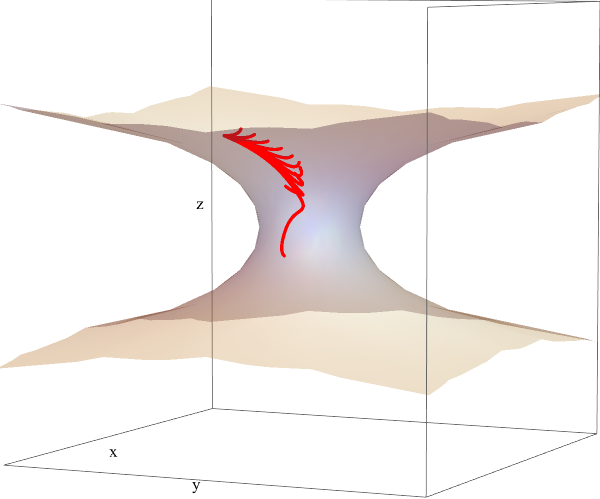}
\caption{Semiclassical trajectories.}
\label{free-multiple-quantum-trajectories}
\end{subfigure}
\caption{ Classical (blue) and effective (red) trajectories with $0\leq a\leq 9$. Semiclassical trajectories get either reflected or transmitted at the throat.}
\label{free-multiple-trajectories}
\end{figure}

We now analyze multiple trajectories by varying $a$ from $(0-9)$, with an incremental step of 1. We obtain classical trajectories in Figure \ref{free-multiple-classical-trajectories} and the corresponding effective trajectories in Figure \ref{free-multiple-quantum-trajectories}. In this case, an even more interesting effect appears for the quantum case: particles with higher energy tunnel through the barrier at the waist; in contrast, the lower energetic ones still get reflected off the barrier.

\begin{figure}
\centering
\begin{subfigure}{.35\textwidth}
  \centering
\includegraphics[width=1\linewidth]{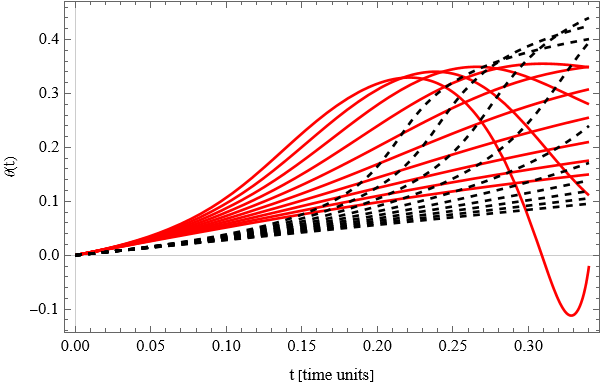}
\caption{$\theta$.}
\label{free-multiple-cvsq-theta}
\end{subfigure}%
  \hfill
\begin{subfigure}{.35\textwidth}
  \centering
\includegraphics[width=1\linewidth]{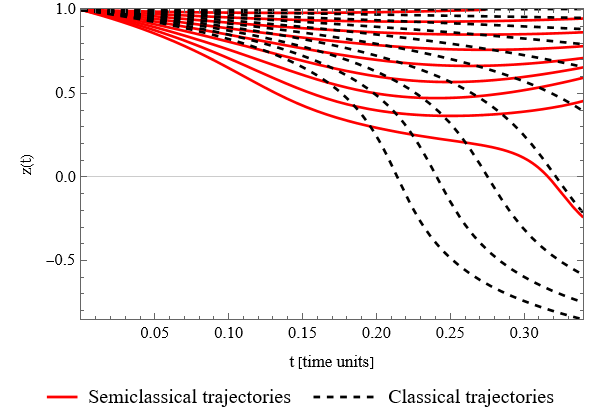}
\caption{$z$ .}
\label{free-multiple-cvsq-z}
\end{subfigure}
\caption{Coordinates evolution for the semiclassical and classical trajectories for free particles with $0\leq a\leq9$.}
\label{fig:test3}
\end{figure}

\begin{figure}
\centering
\begin{subfigure}{.35\textwidth}
  \centering
\includegraphics[width=1\linewidth]{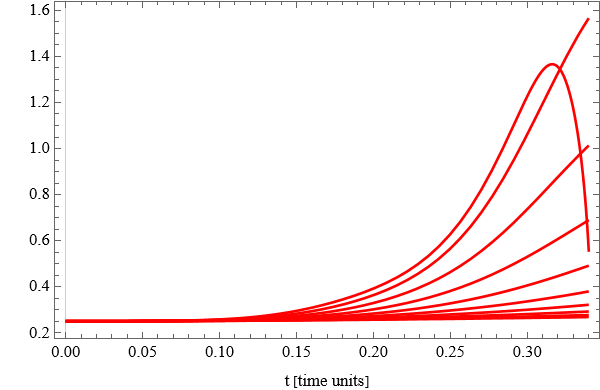}
\caption{$G^{2,0,0,0} G^{0,2,0,0}-(G^{1,1,0,0})^2\geq \hbar^2/4$.}
\label{free-multiple-theta-uncertainty}
\end{subfigure}%
  \hfill
\begin{subfigure}{.35\textwidth}
  \centering
\includegraphics[width=1\linewidth]{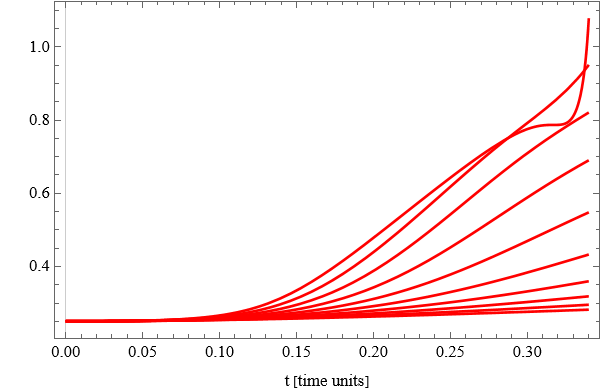}
\caption{$G^{0,0,2,0} G^{0,0,0,2}-(G^{0,0,1,1})^2\geq \hbar^2/4$.}
\label{free-multiple-z-uncertainty}
\end{subfigure}
\caption{Uncertainty relations for each variable with $0\leq a\leq9$.}
\end{figure}

In Figures \ref{free-multiple-cvsq-theta} and \ref{free-multiple-cvsq-z} we compare the $\theta$ and $z$ coordinates for these trajectories, for both the classical and semiclassical solutions. As can be noted, quantum corrections make each variable get more oscillations around its initial value. We also verify, as for the one-particle case, that the uncertainty relation is always valid through the dynamical evolution, as can be appreciated in Figures \ref{free-multiple-theta-uncertainty} and \ref{free-multiple-z-uncertainty} for $\theta$ and $z$ variables, respectively. The trajectories follow from the upper branch to the lower branch of the catenoid.

\subsection{Effective Potential}

We now turn our attention to the effective quantum potential. As our treatment is semiclassical, we obtain the evolution of the system through an extended classical dynamical system, whose qualitative behavior can be determined by analyzing the potential, hence its importance.

The potential can be read off the effective Hamiltonian (\ref{effective_Hamiltonian}), or from its second-order expression (\ref{second_order_Hamiltonian })
\begin{eqnarray}
    V_\text{eff} &=& \frac{\sech^2\left(z/R\right) }{mR^2} \bigg\{-\frac{2\tanh\left(z/R\right)}{R} p_\theta^2 G^{0,1,1,0}+ G^{0,2,0,0}\nonumber\\
    &-&2R\tanh\left(\frac{z}{R}\right) p_\text{z} G^{0,0,1,1} +R^2G^{0,0,0,2} \nonumber\\
    &+&\frac{1}{R^2} \left[2-3\sech^2\left(\frac{z}{R}\right)\right](p_\theta^2+R^2p_\text{z}^2)G^{0,0,2,0}  \bigg\}.
\end{eqnarray}

\begin{figure}[h]
     \centering
     \begin{subfigure}[b]{0.3\textwidth}
         \centering
         \includegraphics[width=\textwidth]{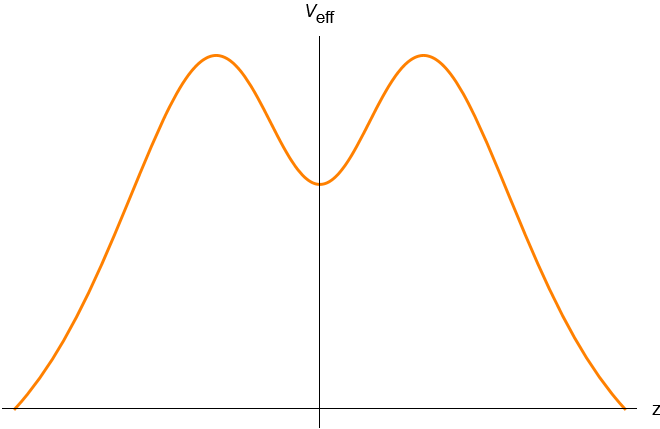}
         \caption{$p_\text{z}=-7.0$}
         \label{fig:veff-b=-7}
     \end{subfigure}
     \hfill
     \begin{subfigure}[b]{0.3\textwidth}
         \centering
         \includegraphics[width=\textwidth]{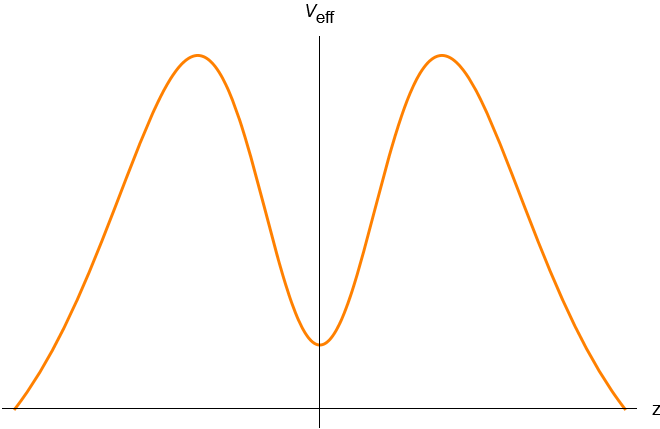}
         \caption{$p_\text{z}=-8.4$}
         \label{fig:veff-b=-8.4}
     \end{subfigure}
     \hfill
     \begin{subfigure}[b]{0.3\textwidth}
         \centering
         \includegraphics[width=\textwidth]{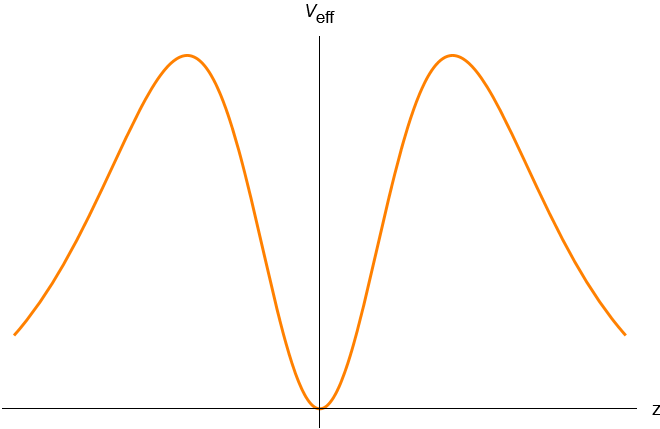}
         \caption{$p_\text{z}=-9.6$}
         \label{fig:veff-b=-9.6}
     \end{subfigure}
        \caption{Classical potential, $V_\text{class}$, for different initial values of $p_\text{z}$.}
        \label{fig:potencial clasico}
\end{figure}

\begin{figure}[h]
     \centering
     \begin{subfigure}[b]{0.3\textwidth}
         \centering
         \includegraphics[width=\textwidth]{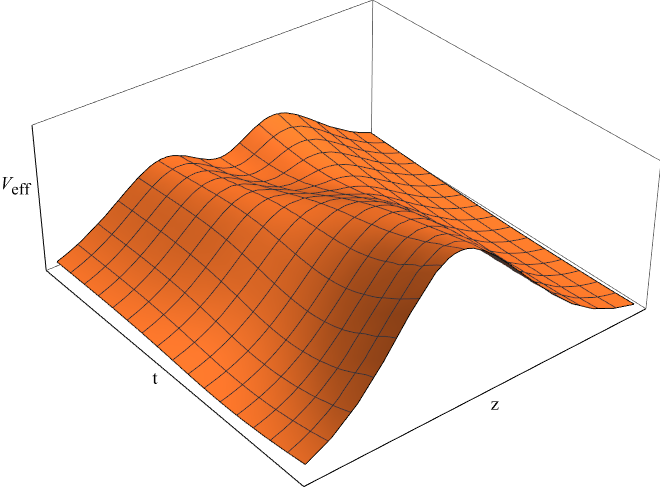}
         \caption{$p_\text{z}=-7.0$}
         \label{fig:veff-b=-7}
     \end{subfigure}
     \hfill
     \begin{subfigure}[b]{0.3\textwidth}
         \centering
         \includegraphics[width=\textwidth]{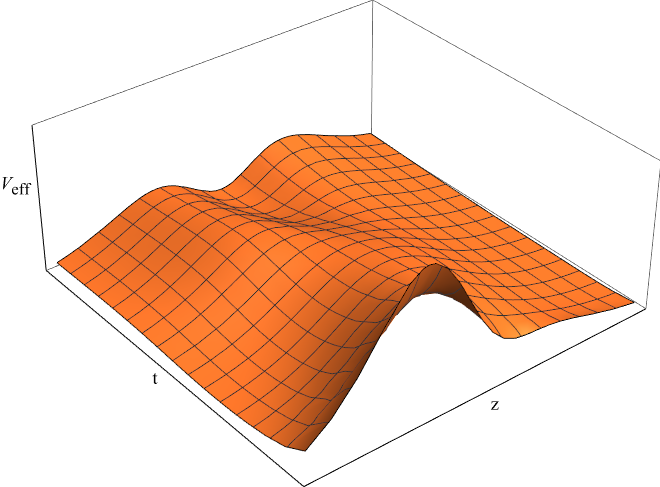}
         \caption{$p_\text{z}=-8.4$}
         \label{fig:veff-b=-8.4}
     \end{subfigure}
     \hfill
     \begin{subfigure}[b]{0.3\textwidth}
         \centering
         \includegraphics[width=\textwidth]{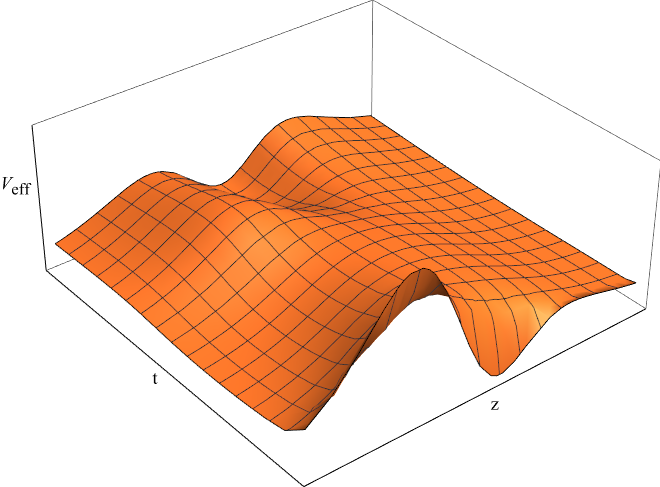}
         \caption{$p_\text{z}=-9.6$}
         \label{fig:veff-b=-9.6}
     \end{subfigure}
        \caption{Effective potential $V_\text{eff}$ for different initial values of $p_\text{z}$.}
        \label{fig:potencial efectivo}
\end{figure}

The classical potential, which in our context, can be interpreted as the initial value of $V_\text{eff}$,
\begin{eqnarray}
    V_\text{class} &=& \frac{\sech^2\left(z/R\right) }{mR^2} \bigg\{ G^{0,2,0,0} _0+R^2G^{0,0,0,2}_0 \nonumber\\
    &+&\frac{1}{R^2} \left[2-3\sech^2\left(\frac{z}{R}\right)\right](p_\theta^2+R^2p_\text{z}^2)G^{0,0,2,0}_0  \bigg\}.\nonumber\\
    ~
\end{eqnarray}
In Figure \ref{fig:potencial clasico} we plot this classical potential for three momentum values, $p_\text{z}=-7.0,-8.4,-9.6$. In contrast, the dynamical effective potential for these same initial values is shown in Figure \ref{fig:potencial efectivo}.

\subsection{Quantum transmission}

We analyze semiclassical trajectories and their passing from the upper branch to the lower branch of the catenoid through the throat at $z=0$, as shown in Figures \ref{free-single-cvsq-trajectory} and \ref{free-multiple-trajectories}. 
In both figures we observe either reflection or transmission off the potential barrier, depending on the initial energy of the particles; such behavior was also obtained in \cite{dandoloff_2010} for example.
We used initial values $p_\text{z}=-7.0, -8.4, -9.6$.

\begin{figure}[h]
     \centering
     \begin{subfigure}[b]{0.3\textwidth}
         \centering
         \includegraphics[width=\textwidth]{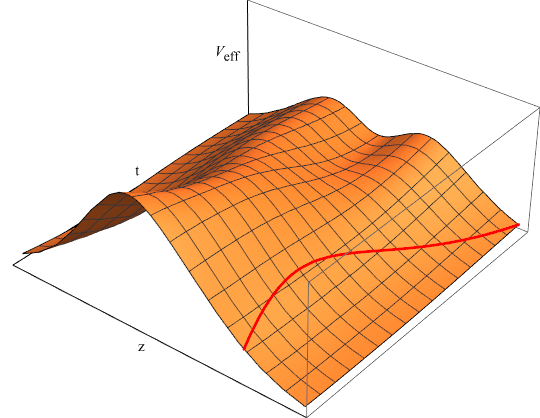}
         \caption{$p_\text{z}=-7.0$}
         \label{fig:semiclassical=-7}
     \end{subfigure}
     \hfill
     \begin{subfigure}[b]{0.3\textwidth}
         \centering
         \includegraphics[width=\textwidth]{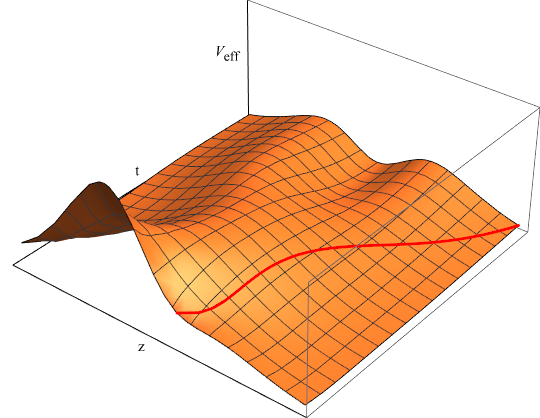}
         \caption{$p_\text{z}=-8.4$}
         \label{fig:semiclassical=-8.4}
     \end{subfigure}
     \hfill
     \begin{subfigure}[b]{0.3\textwidth}
         \centering
         \includegraphics[width=\textwidth]{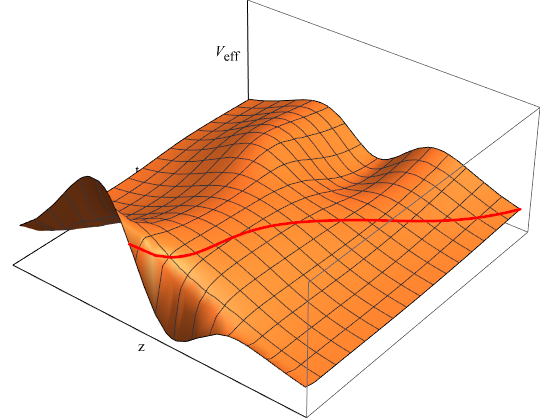}
         \caption{$p_\text{z}=-9.6$}
         \label{fig:semiclassical=-9.6}
     \end{subfigure}
        \caption{Semiclassical trajectories on the effective potential $V_\text{eff}$ for different initial values of $p_\text{z}$.}
        \label{fig:trajectories on potential}
\end{figure}

This can be classically interpreted directly from the effective potential. We can see in figure \ref{fig:trajectories on potential} the  $z$ component of the trajectory atop the effective potential. Initially, the potential presents a local maximum, and no tunneling is possible; the particle gets reflected. As it evolves, the potential changes, acquires local minima, and its maximum decreases, allowing the particle to pass. Whether the particle can tunnel or reflect the potential depends on initial conditions, which, in turn, determine its energy \cite{tunneling}.

\begin{figure}[h!]
     \centering
     \begin{subfigure}[b]{0.3\textwidth}
         \centering
         \includegraphics[width=\textwidth]{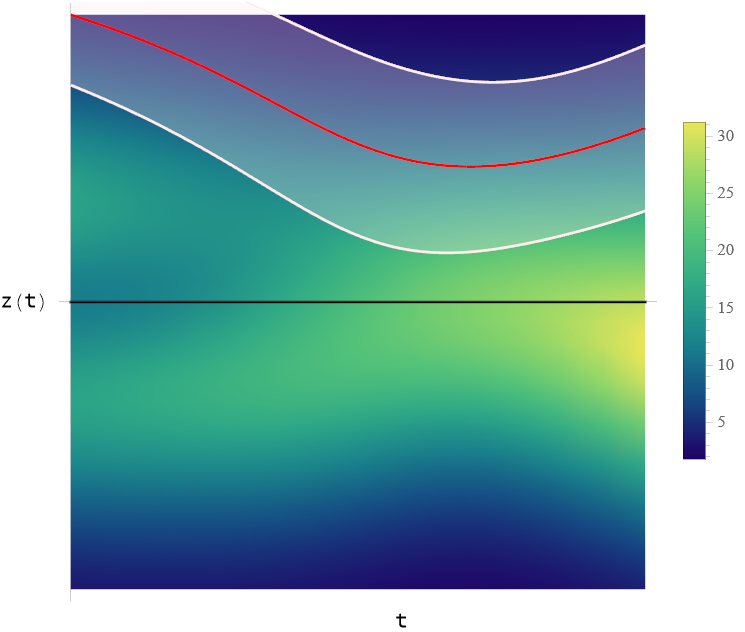}
         \caption{$p_\text{z}=-7.0$}
         \label{fig:tunneling-veff-b=-7}
     \end{subfigure}
     \hfill
     \begin{subfigure}[b]{0.3\textwidth}
         \centering
         \includegraphics[width=\textwidth]{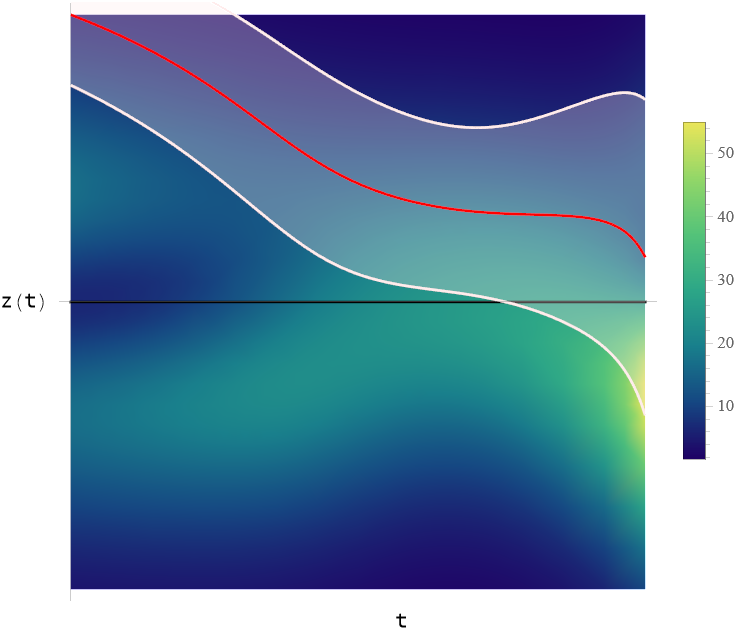}
         \caption{$p_\text{z}=-8.4$}
         \label{fig:tunneling-veff-b=-8.4}
     \end{subfigure}
     \hfill
     \begin{subfigure}[b]{0.3\textwidth}
         \centering
         \includegraphics[width=\textwidth]{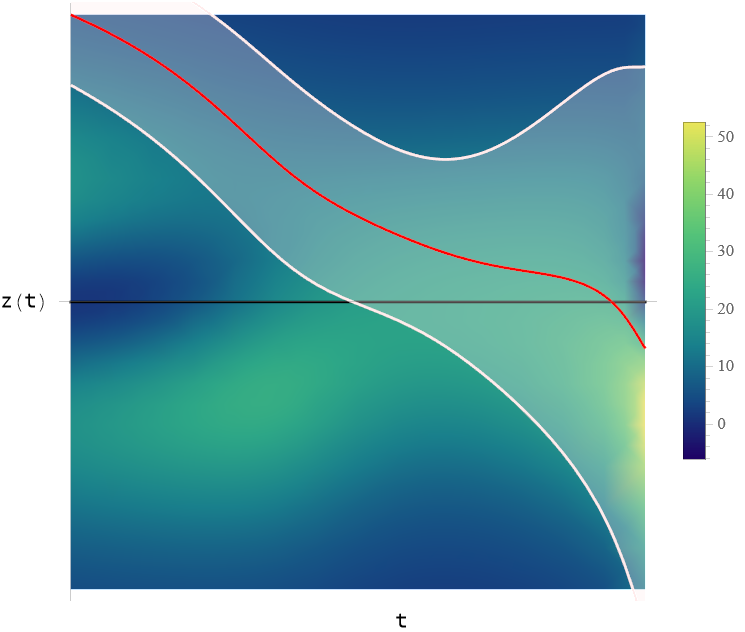}
         \caption{$p_\text{z}=-9.6$}
         \label{fig:tunneling-veff-b=-9.6}
     \end{subfigure}
        \caption{Reflection and transmission effect for semiclassical trajectories with different initial values of $p_\text{z}$.}
        \label{fig:tunneling-catenoid}
\end{figure}

In figure \ref{fig:tunneling-catenoid} we can see the evolution of the $z(t)$ component under the aforementioned initial conditions, with the corresponding effective potential as a density plot. In figure \ref{fig:tunneling-veff-b=-7} the particle gets reflected before being able to reach the throat of the catenoid due to quantum back-reaction. By increasing the value of the initial momentum of the particle $p_z$, the semiclassical particle is able to cross the potential barrier once $p_\text{z}=-9.6$, as we can appreciate in figure \ref{fig:tunneling-veff-b=-9.6}, this is accomplished by the decreasing of the effective potential maxima (figure \ref{fig:semiclassical=-9.6}). Under the momentous quantum mechanics formalism, this has already been interpreted as a quantum tunnelling effect \cite{tunneling,article1,article3,javi}. 

The Supplementary Material presents a GIF animation of the trajectory with initial momentum varying in the range $-9.7 \leq p_\text{z} \leq -7.0$. We can see how the particle crosses the catenoid throat as its initial momentum $p_\text{z}$ increases.

%In the video \ref{vid:tunneling}, we present an animation in GIF format with varying initial momentum in the range $-9.7 \leq p_\text{z} \leq -7.0$. There we can see how the particle crosses the throat of the catenoid as its initial momentum $p_\text{z}$ increases.

%\begin{video}
%\includegraphics[width=0.3\textwidth]{images/videopra.png}
%\url{https://drive.google.com/file/d/1dDosHEPsPMtKmvs_YK8cGyiOH4m0MnUs/view?usp=sharing}
%\caption{\label{vid:tunneling} Particle tunneling through the catenoid throat animation.}
%\end{video}

\section{Discussion} \label{section:discussion}

In this work, we have obtained the effective evolution of a quantum particle constrained to the curved two-dimensional surface of the catenoid. The analysis has been performed with momentous quantum mechanics, a semiclassical description based on expectation values of observables and quantum dispersions.

We have found that the constrained particle exhibits reflection and transmission at the catenoid throat, described in terms of particle trajectories. This same effect has been reported under a different approach in \cite{dandoloff_2010}.
In our case, the dynamics of quantum trajectories do not depend on the quantum state but rather on the extended effective Hamiltonian. 

In the constrained description on a catenoid, one finds that the geometric potential is $ V_\text{G}= - \hbar^2 \sech^4 \left(z/R \right)/2m R^2$,
which maximizes at $z=0$. The presence of this geometric potential leads us to expect a tunneling effect at the surface throat once quantization is considered, i.e., we can find transmission and reflection of quantum particles on the catenoid due to the geometric potential \cite{dandoloff_2010,griffiths,sakurai}. This behavior can be appreciated in the semiclassical trajectories, a particular feature of our approach. This frame is equivalent to the full quantum description.

A very interesting description of the quantum evolution for this constrained system can be given due to the semiclassical nature of momentous quantum mechanics. The classical (confining) potential gets quantum backreacted by quantum dispersion, rendering an effective potential. It is dynamic and evolves with the system: depending on initial conditions for the classical and quantum variables, the potential gives way to the possibility of tunneling, which we depict in our analysis. The occurrence of these effects, inherited from the tunneling effect, has been studied with the effective model of moments in other physical systems. \cite{tunneling}.

We have provided a robust scheme to study several effects in constrained quantum systems, particularly in curved spaces. By extending the conditions discussed in this study to allow more general, or different physical systems, one can explore several models and experimentally interesting settings. We can mention: the transport of particles, a topic of research in vastly different areas such as nanomaterials, radiation physics and materials science \cite{danani}. The catenoid is of particular interest because it is a minimal surface \cite{silva_2021} and minimizes the Willmore energy \cite{willmore}. This makes the surface suitable for making connections between nanostructures of different materials, such as graphene and boron nitride \cite{silva_2021,alsham}, and it has been studied using the Schrödinger formalism \cite{dandoloff_2009,dandoloff_2010}. Such analysis is under further investigation.

\appendix
\section{Symplectic Structure of the Quantum Variables}
\label{appendix-A}

Classical and quantum variables are symplectically orthogonal, which means that their Poisson brackets follow
\begin{equation}
    \bigg\{x_a, G^{a_1,b_1,\cdots,a_k,b_k} \bigg\}=\bigg\{p_a, G^{a_1,b_1,\cdots,a_k,b_k} \bigg\}=0,\label{GX}
\end{equation}
while the Poisson bracket between the quantum variables is non-zero \cite{article1} and is given by, %\ref{https://doi.org/10.48550/arxiv.1110.3337}

\begin{eqnarray}
    &&\bigg\{  G^{a_1,b_1,\cdots,a_k,b_k}, G^{c_1,d_1,\cdots,c_k,d_k}\bigg\}=\nonumber\\
    &&\sum\limits_{i=1}^k \biggl(a_id_i G^{a_1,b_1,\cdots,a_i-1,b_i,\cdots,a_k,b_k} G^{c_1,d_1,\cdots,c_i,d_i-1,\cdots, c_k,d_k\nonumber}\\
    &-&b_i c_i G^{a_1,b_1,\cdots,a_i,b_i-1,\cdots,a_k,b_k} G^{c_1,d_1,\cdots,c_i-1,d_i,\cdots, c_k,d_k}\biggl)\nonumber\\ &+&\sum\limits_n\sum\limits_s\sum\limits_{e_1,\cdots,e_k} (-1)^s \left(\frac{\text{i}\hbar}{2}\right)^{n-1} \delta_{\sum_i e_i,n}\nonumber\\
    &\times& \mathcal{K}^{n,s\{e\}}_{ \{a\},\{b\},\{c\},\{d\} } G^{a_1+c_1-e_1,b_1+d_1-e_1;\cdots; a_k+c_k-e_k,b_k+d_k-e_k},\nonumber\\
   ~ \label{GG}
\end{eqnarray}
where $n=1,\cdots,\tilde{N}$, and 
\begin{equation}
    \tilde{N}=\begin{cases}
    1, & \eta_i\leq 1\\
    \eta_i-1, &\eta_i>1
    \end{cases}
\end{equation}
defining $\eta_i=\sum_i\left(\min[a_i,d_i]+\min[b_i,c_i] \right)$, also, $s=0,\cdots n; 0\leq e_i\leq \min[a_i,d_i,s]+\min[b_i,c_i,n-s]$.\\

On the other side, $\mathcal{K}$ is given by
\begin{eqnarray}
    \mathcal{K}^{n,s\{e\}}_{ \{a\},\{b\},\{c\},\{d\} }&=&\nonumber\\
    \sum\limits_{g_1,\cdots,g_k} \frac{\delta_{\sum_ig_i,n-s}}{s!\left(n-s\right)!}&&\prod\limits_i \frac{\binom{a_i}{e_i-g_i}\binom{b_i}{g_i}\binom{c_i}{g_i}\binom{d_i}{e_i-g_i}}{\binom{n-s}{g_i}\binom{s}{e_i-g_i}},
\end{eqnarray}
with $\max[e_i,s,e_i-a_i,e_i-d_i,0]\leq g_i\leq \min[b_i,c_i,n-s,e_i]$.\\

\section{Canonical Transformation of the Classical Hamiltonian}
\label{appendix-B}
It is interesting to write the classical Hamiltonian (\ref{classical-hamiltonian}) in kinetic plus potential form; this is because there exists an alternate formulation of momentous quantum mechanics in terms of explicitly canonical variables, one for each degree of freedom in the theory \cite{baytas_2020}. This formulation is interesting because there is no need to truncate the Hamiltonian, and one obtains an ``all-order"  potential.

To this end, we perform a canonical transformation to obtain the kinetic terms explicitly in the classical Hamiltonian for the particle on a catenoid (\ref{classical-hamiltonian})
\begin{equation}
    H =\frac{ \sech^2\left(z/R\right) }{2mR^2} \left( p_\theta^2 +R^2p_\text{z}^2 \right). \nonumber
\end{equation}
We propose the following redefinition
\begin{equation}
    P_2= \sech\left(\frac{z}{R}\right)p_\theta,
\end{equation}
and we use the second canonical generating function \cite{goldstein:mechanics},
\begin{equation}
    \frac{\partial F_2}{\partial \theta} = p_\theta.
\end{equation}
We get
\begin{equation}
    F_2 = \theta \cosh\left(\frac{z}{R}\right) P_2 +g(z) P_1,
\end{equation}
where we have introduced the second term on the r.h.s. $g(z)P_1$, such that $F_2=\sum\limits_{i=1}^2 Q_i P_i$, to have the canonical pairs $\left(Q_i,P_i \right)$. Taking the derivate with respect to $z$, we identify,
\begin{equation}
    \frac{\partial F_2}{\partial z} = \frac{\theta}{R}\sinh\left(\frac{z}{R}\right) P_2+g ^\prime (z) P_1 = p_\text{z}.
\end{equation}
Now we propose that $g^\prime (z) = \cosh\left(\frac{z}{R}\right)$, such that the generating function becomes 
\begin{equation}
    F_2=R\sinh\left(\frac{z}{R}\right) P_1+ \theta \cosh\left(\frac{z}{R}\right)P_2 .
\end{equation}
From the generating function, we identify the new canonical pairs of the transformation in terms of the old variables, which are, respectively,
\begin{eqnarray}
    Q_1 &=& R\sinh\left(\frac{z}{R}\right),\nonumber\\
    P_1 &=& \sech\left(\frac{z}{R}\right) p_\text{z} -\frac{\theta}{R} \sinh\left(\frac{z}{R}\right) \sech^2\left(\frac{z}{R}\right) p_\theta,\nonumber\\
    Q_2 &=&\theta \cosh\left(\frac{z}{R}\right),\nonumber\\
    P_2 &=&\sech\left(\frac{z}{R}\right) p_\theta,
\end{eqnarray}
 and $\left\{Q_i,P_j\right\}=\delta_{i,j}$. Upon substitution, they allow us to rewrite the Hamiltonian $H(q,p,t)$ as a new Hamiltonian $K(Q,P,t)$, given $\partial F_2/\partial t=0$. The new Hamiltonian is,
\begin{equation}
    K(Q,P,t)= \frac{P_1^2}{2m} + \frac{P_2^2}{2mR^2}+\frac{Q_1Q_2 P_1 P_2}{m \left(R^2+Q_1^2 \right)} + \frac{Q_1^2 Q_2^2 P_2^2}{2m \left(R^2+Q_1^2 \right)^2}. \label{Kamiltonian}
\end{equation}

We can see from (\ref{Kamiltonian}), that the Hamiltonian (\ref{classical-hamiltonian}) corresponds to a constrained Hamiltonian in the usual way, as the first two terms in $K(Q,P,t)$ are kinetic terms, while the last two terms correspond to a constraining potential, which depends on position and momentum. An extension of the present work in this canonical formulation will be implemented in a future analysis.

\bigbreak

%\printbibliography
%\section*{References}
%\begin{enumerate}[label=\text{[}\arabic*\text{]}]

\end{document}